\documentclass{article}

\usepackage{arxiv}

\usepackage[T1]{fontenc}    
\usepackage{hyperref}       
\usepackage{url}            
\usepackage{booktabs}       
\usepackage{amssymb,amsmath,amsthm, amsfonts}
\usepackage{nicefrac}       
\usepackage{microtype}      
\usepackage{lipsum}
\usepackage{listings}
\usepackage{graphicx}
\usepackage{xcolor}
\usepackage{setspace}
\definecolor{light-gray}{gray}{0.97}
\usepackage[noend]{algpseudocode}

\newtheorem{theorem}{Theorem}

\begin{document}
	
\title{Fairness through Experimentation: Inequality in A/B testing as an approach to responsible design.}
\newcommand{\runtitle}{Fairness Through Experimentation}
\newcommand{\li}{LinkedIn Corporation \\ Sunnyvale, CA \\ United States of America\\}
\author{
	Guillaume Saint-Jacques \\
	\li
	\texttt{gsaintjacques@linkedin.com} \\
\And
	Amir Sepehri  \\
	\li
	\texttt{asepehri@linkedin.com} \\
	\And
	Nicole Li  \\
	\li
	\texttt{nili@linkedin.com} \\
	\And
	Igor Perisic  \\
	\li
	\texttt{iperisic@linkedin.com} \\
}

\maketitle
	
\begin{abstract}
	As technology continues to advance, there is increasing concern about individuals being left behind. Many businesses are striving to adopt responsible design practices and avoid any unintended consequences of their products and services, ranging from privacy vulnerabilities to algorithmic bias. 
	
	We propose a novel approach to fairness and inclusiveness based on experimentation. We use experimentation because we want to assess not only the intrinsic properties of products and algorithms but also their impact on people. We do this by introducing an inequality approach to A/B testing, leveraging the Atkinson index from the economics literature. We show how to perform causal inference over this inequality measure. We also introduce the concept of site-wide inequality impact, which captures the inclusiveness impact of targeting specific subpopulations for experiments, and show how to conduct statistical inference on this impact. 
	
	We provide real examples from LinkedIn, as well as an open-source, highly scalable implementation of the computation of the Atkinson index and its variance in Spark/Scala. 
	
	We also provide over a year's worth of learnings -- gathered by deploying our method at scale and analyzing thousands of experiments -- on which areas and which kinds of product innovations seem to inherently foster fairness through inclusiveness.
\end{abstract}

\textbf{Acknowledgements:} 
We would like to thank Meg Garlinghouse, Ya Xu, Stuart Ambler, James Eric Sorenson, Bari Lemberger, 
Parvez Ahammad, Sofus Mack\'{a}ssy, Steve Lynch, as well as the entire LinkedIn Data Science Applied Research team, and the Computational Social Science team for valuable contributions and comments.

\tableofcontents

\spacing{1.5}

\section{Introduction}

\subsubsection*{Motivation}

Businesses are increasingly striving to adopt responsible design
practices to avoid unintended consequences of their products and
services. Examples range privacy vulnerabilities \cite{Cyphers2019} to claims of algorithmic bias
against minorities \cite{Schuetz2019}, such as women and ethnic minorities \cite{Hao2019}.
 A growing literature on algorithmic
fairness \cite{agarwal2018reductions,dwork2012fairness,barocas2017fairness,bolukbasi2016man} focuses on guaranteeing that algorithms provide equal
representation of various demographics, or that their error rates are
consistent across different demographics.~

However, focusing exclusively on algorithms and their properties can
lead one to miss two crucial points:

First, how people react and interact with a product is as important as
the product itself. Instead of only focusing on whether
an~\emph{algorithm}~is representative or fair, we should also focus on
the~\emph{outcomes}. For example, an algorithm may seem to treat men and
women similarly, but if it results in women getting
disengaged, it may be an undesirable outcome.

Second, the focus on protected categories can be too narrow, especially
in the age of the internet. In particular, many products now have a
``network'' or ``social'' component and a user's experience may be determined by who they know: ~\emph{share this with your friends,
post this to your community, refer your friends for employment
opportunities\ldots{}} In other words, many features often risk increasing the gap between people with strong networks and people without them. While ``network strength'' is not a legally protected category, this is still something worth paying attention to. Worse, how many more such categories may we all be overlooking? We need a way to be alerted if a feature we build has an inequality impact, even if it is not between categories we are explicitly monitoring.

Therefore, even if a product may seem to have been designed in a
``responsible'' or ``fair'' manner, it can still drive a wedge between
different groups of users, in particular groups of users that a company
may not be explicitly monitoring (for example, between people who have
a powerful social network, and those who do not).~

\subsubsection*{An A/B testing approach to fairness}

We propose a novel approach to help address these concerns. We use it as
a complement, not a substitute, to more traditional algorithmic bias
approaches. We leverage experimentation, which, instead of
characterizing a feature in the void, measures the effect it has on
users. In particular, we measure~\emph{inequality impact}~of all new
features and product changes (whether they are algorithms, UI changes, or infrastructure changes, for example) and flag experiments that have a notably
positive or negative inequality impact. In essence, along with asking,
"what would be the total number of sessions or contributions on our
platform" if feature A (vs. B) were rolled out?, we also ask "if feature A
were to be rolled out, what would be the share of the top 1\%, in terms
of engagement and contributions? Would inequality go up or down?"
This combination of concepts from measures of inequality, drawn from the
economics literature, with A/B testing provides us two distinct
advantages.

First, instead of just~\emph{measuring}~inequality we can trace it back
to its~\emph{causes}: a specific set of features and product decisions.
We can ask: which of our products is making the 'rich' richer?~

Second, compared to classical algorithmic fairness approaches, it also
gives us another advantage: it helps us identify features that increase
inequality without having to rely only on explicitly protected categories. A feature
that increases the gap between any two classes of users is likely to be
detected.

We then apply this method to every single experiment that LinkedIn has
conducted over the past year and analyze the ones with the highest
inequality-increasing or inequality-reducing impact.~

The remainder of this paper is organized as follows: First, we provide
an introduction to our inequality impact measurement method, relying on an
adequately parametrized Atkinson index. Second, we show how this method
can be combined with A/B testing, comparing a treatment and a control
group via a difference in Atkinson indices. We discuss the variance of
this difference and statistical inference, as well as extrapolation to
the effect beyond the experiment. Third, we describe our process for
driving large-scale business impact using this method, discussing
scalable implementation as well as organizational processes needed.
Fourth, we share a specific example of a LinkedIn experiment that was
identified by this method as having a strong inequality impact. 
Finally, using the thousands of experiments we have analysed, we share our high-level findings of the
types of interventions that generally seem to be inequality-increasing
and decreasing, as well as the best use cases for this approach.~

\section{The Atkinson Index for fairness}

We start by describing the Atkinson index as a measure of inequality,
its desirable properties, as well as how to parametrize it.

\subsection{A highly stylized example}

The most common way of characterizing a dataset is using its average.
Most experimentation platforms primarily report the~\emph{average
	effect}~of a new feature for a given population segment. They compute the average of business metrics
for users who were exposed to the new feature (often
called~\emph{treatment group}), and compares them to the average for
members who were not (\emph{control group).~}The main benefit of this
method is its simplicity. It also easily allows us to extrapolate
the~\emph{total}~effect on business metrics if the new product were to be given
to all users (i.e.,~\emph{our website has 1M sessions today; it would
	have 1.05M sessions if we rolled this feature out).~}

\begin{figure}
	\begin{center}
		\includegraphics[width=0.8\textwidth]{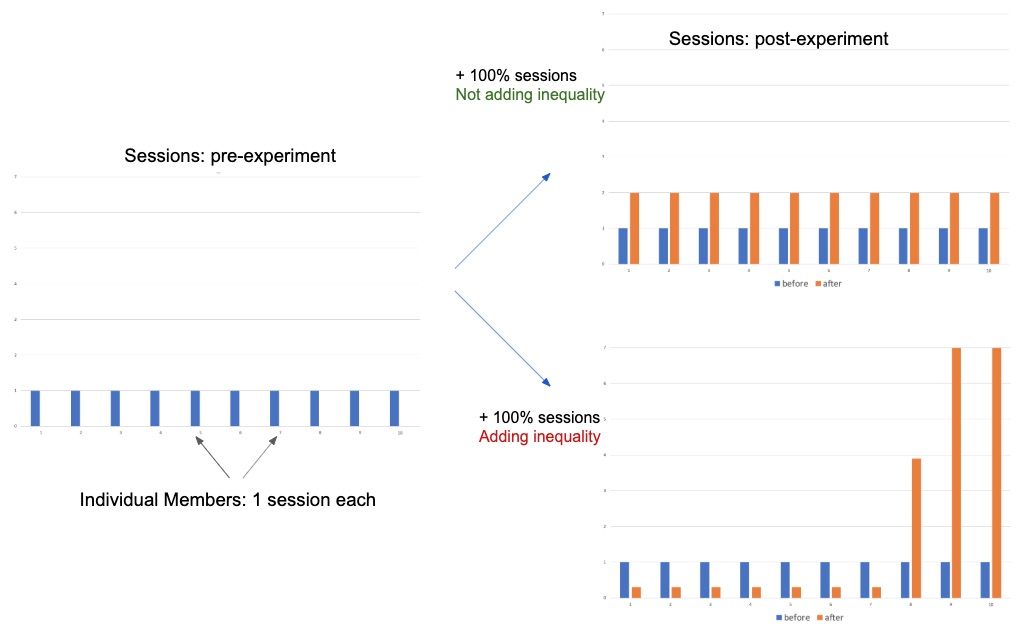}
		\caption{Two products with the same average impact, but different individual effect patterns.} \label{atk-example}
	\end{center}
\end{figure}

However, exclusively focusing on the average effect may hide significant
disparities between users. In particular, it could conceal significant
disparities that are being introduced by the new feature. For example,
imagine a website that only has ten users, which all visit the site once
a day (1 session/user). We depict this on the left side of Figure
\ref{atk-example}. An initiative that doubles sessions
could do so in many ways: it could either double every user's sessions
(top of the figure), or more than quadruple some user's sessions and
decrease others (bottom). Yet that difference is typically not readily
visible on traditional platforms, which only show averages. If a company
needs to tell these two types of scenarios apart, then a method is
required in order to automatically detect interventions that may
increase inequality. This is the case for LinkedIn, whose vision is to create
economic opportunity for every member of the global workforce.

\subsection{Properties of the Atkinson Index}

First introduced by Atkinson \cite{atkinson1970}, the
Atkinson index is a standard measure of inequality that economists
frequently use for country-to-country comparisons of income distributions \cite{atkinson2010analyzing}. 
Given a sample $x_1, \ldots,x_n$, representing a metric (such as sessions, page views, job applications, \ldots) for a group of users indexed by \emph{i}, and
an inequality-aversion parameter $\epsilon$. It is computed as follows for $\epsilon \neq 1$:

\begin{equation} \label{atk}
A_\epsilon (x_1, \ldots, x_n) = 1 - \frac{(\frac{1}{n} \displaystyle \sum_{i=1}^n x_i^{1-\epsilon})^\frac{1}{1-\epsilon}}{\frac{1}{n}\displaystyle \sum_{i=1}^n x_i}.
\end{equation}

Intuitively, for a fixed $\epsilon$ , a low value of $A_\epsilon$ is obtained when all individuals have almost the exact same metric value. When this is exact, the Atkinson index is zero. A higher value reflects more inequality. The $\epsilon$ parameter can be tuned to reflect the inequality preferences of a decision maker, as will be shown in the next section.

That Atkinson index has several desirable properties
\cite{atkinson1970} for our application, namely:

\emph{It is non-negative and is equal to zero only if all individual
	metrics are the sam}e. This is helpful in comparing distributions to the
``pure equality'' baseline, as shown in the above figure.

\emph{It satisfies population replication axiom}: if we create a new
population by replicating the existing population any number of times,
the inequality remains the same. This is particularly helpful because
companies tend to grow, and the simple fact that the number of users is
growing should not change the inequality index, unless actual inequality
increases are happening at the same time.

\emph{It satisfies the principle of transfers}: any redistribution
(i.e., reducing a user's metrics to increase the metric of another,
lower-ranking user) results in a decrease of the Atkinson index, as long
as this intervention does not change their rank in the distribution. This can be useful to
specifically assess the impact of ``redistribution'' experiments, such
as redistributing attention on the feed
\cite{simpsonpaper}.

\emph{If all metrics are multiplied by a positive constant, the index
	remains the same}: this is useful because it allows us to compare
inequality of distributions across different time horizons meaningfully.
For example, if all users have the same number of sessions every day,
the inequality will be the same, whether measured as daily, weekly, or
monthly sessions.

\emph{It is subgroup decomposable}: Any Atkinson index can be decomposed as an average of indices for subpopulations. This is useful when an experiment, for example, only targets a specific population, but we would like to understand its inequality effect on the population as a whole. \cite{das1982decomposition, lasso2003casilda}

\emph{It is parametrizable:}~different levels of inequality may be
acceptable for different decision-makers, different use cases or metrics, and we can parametrize
inequality aversion using $\epsilon$.

\emph{It is scalable:~}
Looking at the formula above, one can see that
the Atkinson index lends itself to distributed computation, using
map-reduce or Spark, for example.

\subsection{Parametrizing the Atkinson Index (choosing $\epsilon$)}

\begin{figure}
	\begin{center}
		\includegraphics[width=0.8\textwidth]{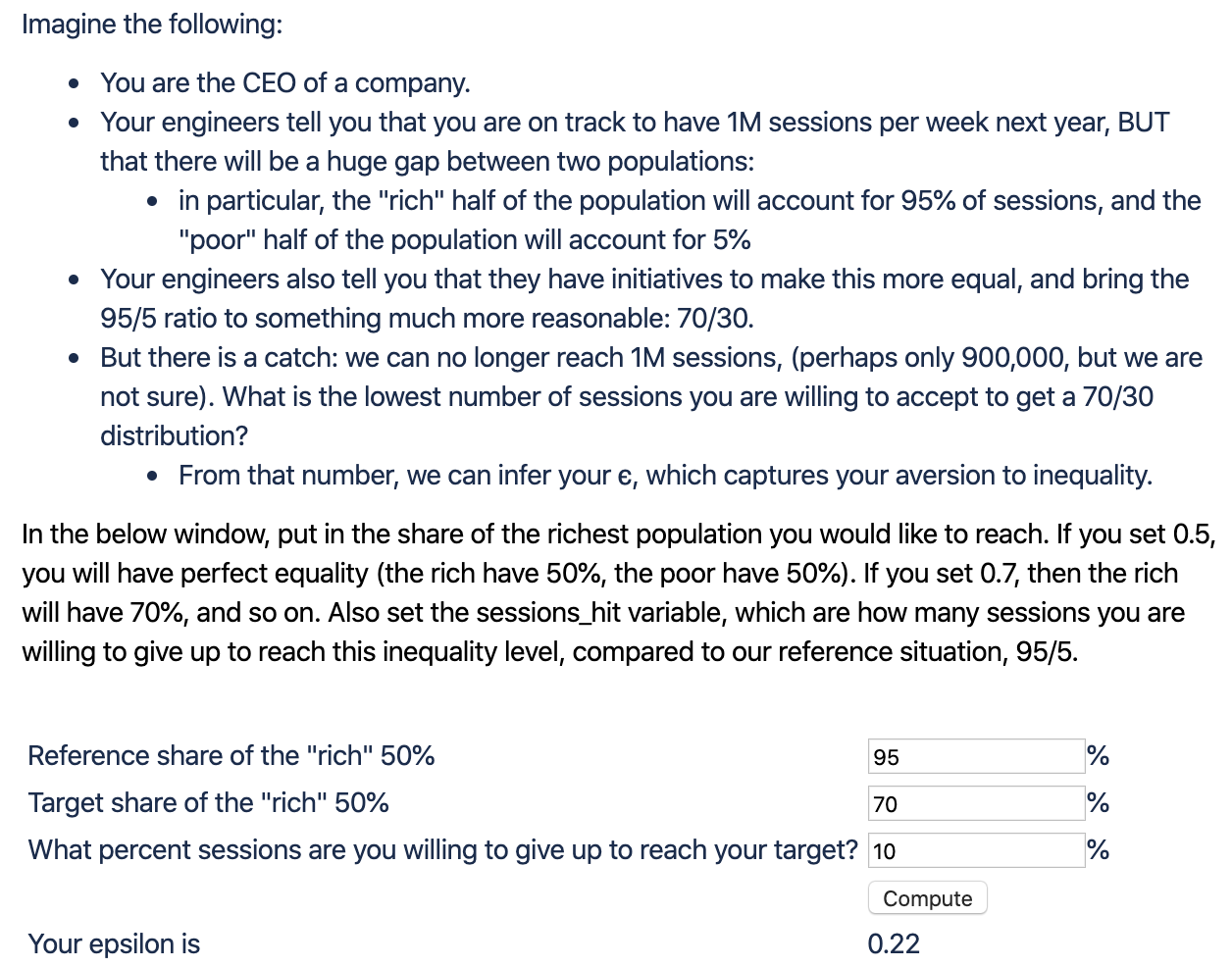}
		\caption{Screenshot of a choice experiment to elicit $\epsilon$} \label{choiceExp}
	\end{center}
\end{figure}

The only parameter for the Atkinson index is epsilon, which measures
a decision-maker's inequality aversion. A good way to parametrize this is
through choice experiments: presenting individuals with a series of
choices between two possible simple distributions, and asking them which
one they prefer. In particular, we simulate cases where reducing
inequality is possible, but at a cost: the total of the metric must be
reduced (see figure \ref{choiceExp} for an illustration). Eliciting the acceptable cost level maps to the decision
maker's epsilon.

Following \cite{creedy2016}, we can model the Atkinson
index as a discount factor for the total of the metric. If $U$ is the
utility of the decision-maker, we have

\begin{equation}\label{utility}
U(T,D,\epsilon) = T  (1-A(\epsilon,D))
\end{equation}

T is the total of the metric, and $D$ is the distribution. Notice that if
there is no inequality, the inequality index is zero, and the decision
maker's utility is equal to the total of the metric. In other words, for
each level of inequality, there is an equally distributed equivalent
total, which the decision-maker would consider equivalent.

In particular, we simulate simple, two-individual cases. In the two-individual case,
 the Atkinson index can be written as follows:

\begin{displaymath}
A(y_{1}, y_{2}, \epsilon) =
-\frac{2}
{{\left(y_{1} + y_{2}\right)} {\left(\frac{1}{2} \, y_{1}^{-\epsilon + 1} + \frac{1}{2} \, y_{2}^{-\epsilon + 1}\right)}^{\left(\frac{1}{\epsilon - 1}\right)}} + 1
\end{displaymath}

Equivalently, it can be rewritten as a function of the total and the
share of the richest individual:
{ 
	\begin{multline}\label{equivalence_atk}
	A_S(T, s_{1}, \epsilon) = \\
	\frac{2}
	{{\left(T {\left(s_{1} - 1\right)} - T s_{1}\right)} {\left(\frac{1}{2} \, \left(-T {\left(s_{1} - 1\right)}\right)^{-\epsilon + 1} + \frac{1}{2} \, \left(T s_{1}\right)^{-\epsilon + 1}\right)}^{\left(\frac{1}{\epsilon- 1}\right)}} + 1
	\end{multline}
}

We use this fact for our choice experiments: If decision makers truly regard high levels of inequality as problematic, it follow that they should be willing to sacrifice some amount of
core business metrics (say, sessions), in order to see inequality be reduced. Therefore, we provide an individual with a prospective total $T_1$, a prospective share of the richest
$s_1$, and provide an alternative $s_2$, leading to less
inequality. We then elicit the the total $T_2$ that would make these
two scenarios equivalent, which leads us to compute the individual's
$\epsilon$, as a solution to:

\begin{displaymath}
(1-A_S(T_1,S_1,\epsilon)) = (1-A_S(T_2,S_2,\epsilon))\frac{T_2}{T_1}
\end{displaymath}

Figure \ref{choiceExp} illustrates an example of a choice experiment.

\subsection{Interpreting an Atkinson index}

Once the epsilon has been set, we can interpret the Atkinson
index as a discount factor on the total of the metric according to the
individual's preferences, as shown in equation \ref{utility}. In short, as outlined in the previous section, it measures how much of a core business metric (say, sessions) one would be willing to give up in order to see that metric be equally distributed between members. 

Translating this core interpretation into an A/B testing context, this implies that when we see a +0.01 Atkinson impact of an experiment, we can read this as "\textit{someone with my inequality preferences  (i.e., my epsilon) would rather give up 1\% sessions than have this feature be rolled out}". This simple interpretation is quite helpful when interpreting the magnitude of an inequality impact.

\section{Inequality in A/B testing}

We now formally discuss how to leverage the Atkinson Index in an A/B
testing context. In an online experiment, we can measure the Atkinson index for the treatment group and the control group and decide whether they are significantly different. This provides a way to determine whether the treatment product increases inequality amongst users.

\subsection{Difference in means}

Consider an A/B test with two variants, $A$ and $B$, of the product. To compare the relative inequality impact of the two versions, we compute the Atkinson index for both; denote them by \(A_\epsilon^A\) and \(A_\epsilon^B\). We then compare \(A_\epsilon^A\) and \(A_\epsilon^B\) to determine if there is a statistically significant difference between them. If we know the distribution of \(A_\epsilon^A\) and \(A_\epsilon^B\), we can use the  standard Neyman-Pearson \cite{neyman1933ix} hypothesis testing framework to test the null hypothesis that the two variants are similar in terms of inequality impact. That is,

\[H_0: A_\epsilon^A = A_\epsilon^B.\]

To measure the \emph{inequality} impact of an experiment, we use a
difference in means between treatment and control:

$$
\hat{\delta} =
A_\epsilon^T -
A_\epsilon^C
$$

This can simply be read as the inequality increase or decrease that the
new feature introduces, and can be interpreted as outlined above.

\subsection{Variance and Inference}

In order to decide whether an observed difference in inequality between
treatment and control is \emph{significant}, we need to compare it to
its expected variance under the null hypothesis that there is no difference between the two.

\subsubsection*{Variance of one Atkinson index}

To carry out inference, it is crucial to have an approximation of the distribution of $A_\epsilon$. The distribution of the Atkinson index can be approximated using the central limit theorem and the delta method. The following theorem gives the asymptotic distribution.

\begin{theorem}
	For a sample $x_1, \ldots,x_n$, independently distributed according to a distribution $F$, and a fixed aversion parameter $\epsilon$, the Atkinson index $A_\epsilon^{(n)}$ is asymptotically normally distributed. Specifically, the following holds
	\begin{align*}
	\sqrt{n} \big( \frac{A_\epsilon^{(n)} - A_\epsilon(F)}{\sigma_n} \big) \rightarrow N(0,1),
	\end{align*}
	where
	\begin{align*}
	A_\epsilon(F) = 1 - \frac{\left(\int x^{1-\epsilon} dF(x) \right)^{\frac{1}{1-\epsilon}}}{\int x dF(x)}
	\end{align*}
	and
	\begin{equation}
	\begin{split}
	\sigma_n^2 = \widehat{\Sigma}_{11} \times\frac{(\frac{1}{n}\sum_{i=1}^n x_i^{1-\epsilon})^\frac{2 \epsilon}{1-\epsilon}}{(1-\epsilon)^2\left(\frac{1}{n}\sum_{i=1}^n x_i \right)^2 } - 2 \widehat{\Sigma}_{12} \times  \frac{(\frac{1}{n}\sum_{i=1}^n x_i^{1-\epsilon})^\frac{1 + \epsilon}{1-\epsilon}}{(1-\epsilon)\left(\frac{1}{n}\sum_{i=1}^n x_i \right)^3 } + \\ \widehat{\Sigma}_{22} \times \frac{(\frac{1}{n}\sum_{i=1}^n x_i^{1-\epsilon})^\frac{2}{1-\epsilon}}{\left(\frac{1}{n}\sum_{i=1}^n x_i \right)^4 }
	\end{split}
	\end{equation}
	for
	\begin{align*}
	\widehat{\Sigma}_{11} &= \frac{1}{n}\sum_{i=1}^n x_i^{2-2\epsilon} - (\frac{1}{n}\sum_{i=1}^n x_i^{1-\epsilon})^2, \\
	\widehat{\Sigma}_{21} &=  \widehat{\Sigma}_{12} = \frac{1}{n}\sum_{i=1}^n x_i^{2-\epsilon} -(\frac{1}{n}\sum_{i=1}^n x_i)\times (\frac{1}{n}\sum_{i=1}^n x_i^{1-\epsilon}),\\
	\widehat{\Sigma}_{22} &=  \frac{1}{n}\sum_{i=1}^n x_i^{2} - (\frac{1}{n}\sum_{i=1}^n x_i)^2.
	\end{align*}
\end{theorem}
The theorem is proved in the appendix.

\subsubsection*{Variance of the difference between treatment and control}

$$
\hat{V_\delta} = \frac{\sigma^2_{n_T}}{n_T} +
 \frac{\sigma^2_{n_C}}{n_C}
$$

$$
T = \hat{\delta} /
\hat{V_\delta}
$$

We then compare T to the critical values of a normal distribution.

\subsection{Comparison to bootstrap variance}

\begin{figure}
\centering
		\includegraphics[width=0.8\textwidth]{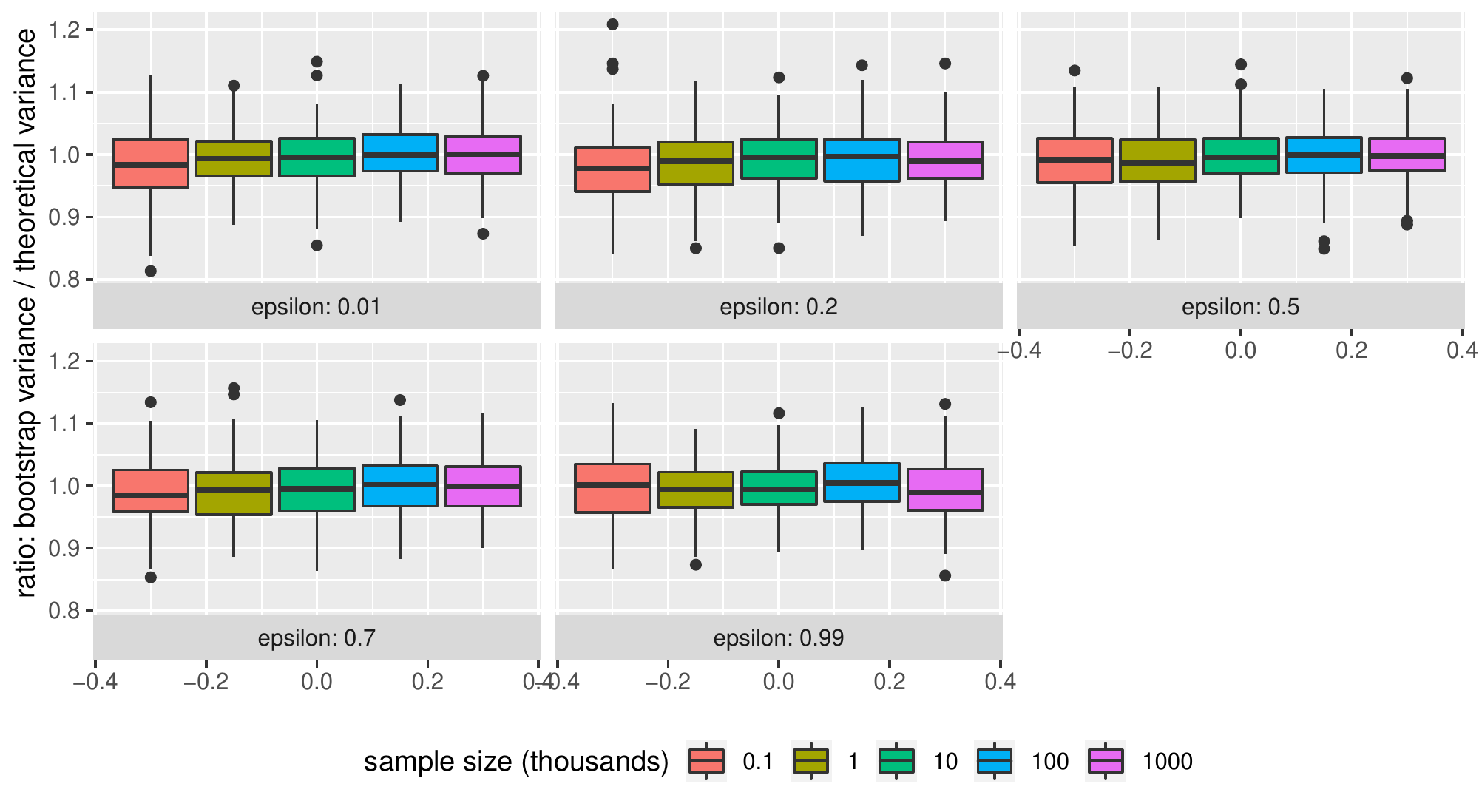}
		\caption{Ratio of 1000-run bootstrap variance to theoretical variance, by $\epsilon$ and sample size. This exercise was repeated 200 times, resulting in a distribution of ratios showed in the boxplots}\label{bootstrap-plots}
\end{figure}

Theorem 1 provides the asymptotic variance to the Atkinson index. For a finite-sample perspective, we compared it to bootstrap simulations on real LinkedIn metrics. The bootstrap variance is typically within $\pm 4\% $ of the theoretical asymptotic variance. 
Table  \ref{tab:bootstrap-variance-table} below shows
a simulation on log-normally distributed synthetic metric data, with different sample
sizes and values of $\epsilon$. All bootstrap variances are estimated using 1000 runs, and then further averaged over 200 runs. We show the variability over these 200 runs in figure \ref{bootstrap-plots}.
\begin{table}[t]

\caption{\label{tab:bootstrap-variance-table}Ratio of bootstrap variance to theoretical variance, by sample size and Atkinson epsilon.
      Bootstrap variance is computed over 1000 runs; these numbers are further averaged over 200 runs.}
\centering
\begin{tabular}{r|r|r|r|r|r}
\hline
N & 0.01 & 0.2 & 0.5 & 0.7 & 0.99\\
\hline
1e+02 & 0.950 & 0.951 & 0.969 & 0.972 & 0.985\\
\hline
1e+03 & 0.983 & 0.979 & 0.983 & 0.991 & 0.991\\
\hline
1e+04 & 0.993 & 0.996 & 0.997 & 0.996 & 0.997\\
\hline
1e+05 & 1.004 & 0.994 & 0.999 & 1.003 & 1.006\\
\hline
1e+06 & 1.001 & 0.993 & 0.999 & 0.999 & 0.992\\
\hline
\end{tabular}
\end{table}

\subsection{Extrapolating to company-wide inequality impact}\label{swi-atk}

A standard analysis of treatment impact on inequality compares the Atkinson index in the treatment and control group, and whether the difference is statistically significant. This indeed is a crucial piece to understanding the inequality impact of the treatment. However, it does not capture the full picture. Imagine our member base can be decomposed into three groups. Group A has very little engagement. Group B has more, and group C is the most highly engaged. Now consider an experiment that targets only groups B and C, leaving A out of the experiment completely. Imagine that this experiment is only successful at increasing engagement for group B, and has no actual impact on group C. This treatment reduces inequality \textit{within} the experiment population (B+C):  the engagement of B went up, and the engagement of C remained constant.  However, if we consider the whole member base (A+B+C), treatment probably increases engagement inequality, as it increases the gap between the already engaged users (B+C ) and the others(A). Therefore, we also need to consider the impact of the treatment on inequality in the broader member base, rather than only on the population targeted by the experiment.

\subsection{Site-wide estimation of the Atkinson index}
One way to measure the broader impact on inequality is to estimate the change in Atkinson index if we were to fully ramp the treatment (100\% ramp) to the segment of members the experiment is targeting.

\begin{figure}
	\begin{center}
		\includegraphics[width=0.8\textwidth]{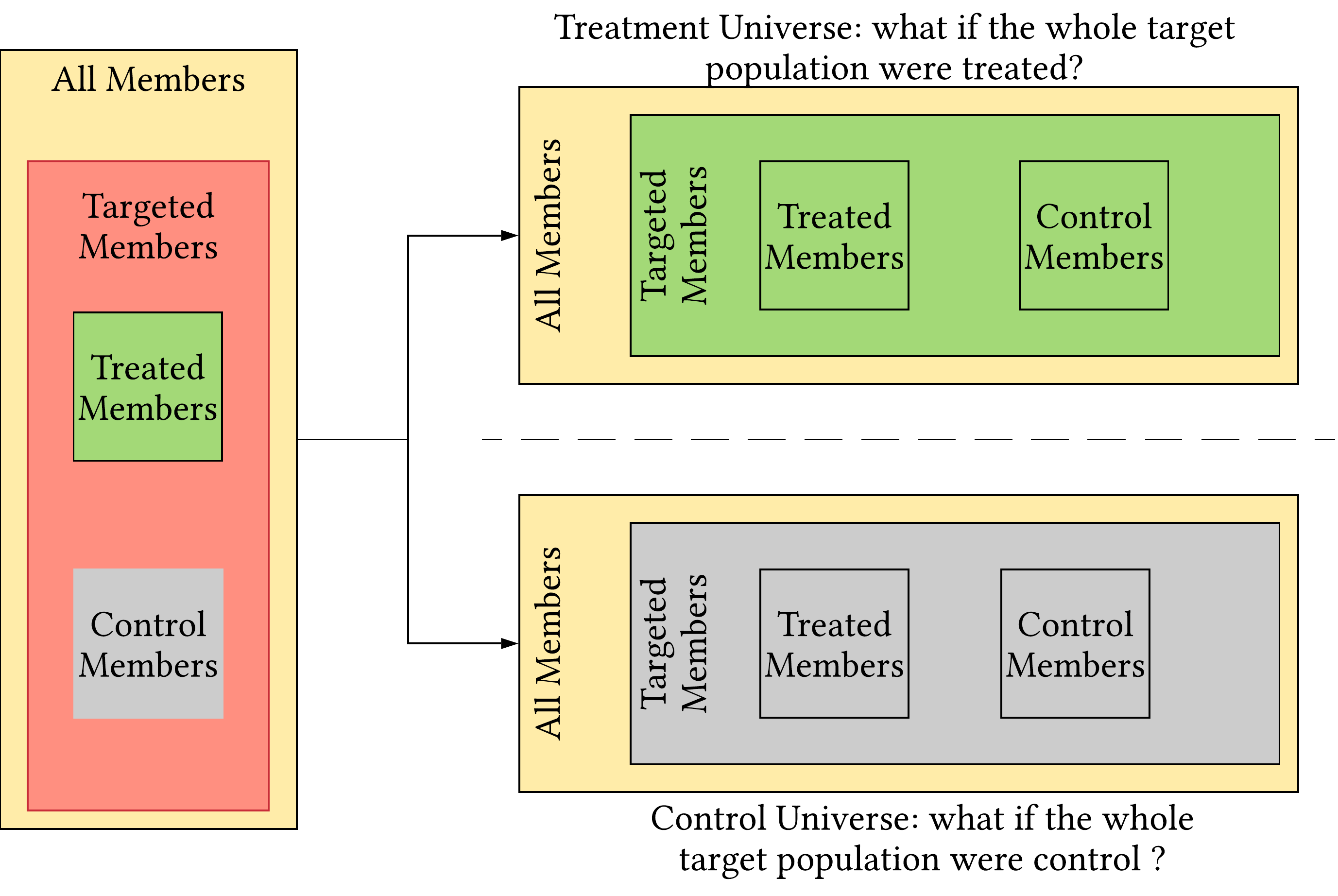}
		\caption{Two universes before and after releasing a product.} \label{SWA}
	\end{center}
\end{figure}

Hypothetically, assume that we were to ramp treatment to 100\% in the segment it's targeting. We would like to compare this scenario to the case where we ramp control to 100\% in the same segment. The two hypothetical scenarios are illustrated in Figure \ref{SWA}, where each color corresponds to a variant.

To estimate the impact in each of the two scenarios, we can extrapolate the impact in the targeted segment based on the A/B results, and use the existing results for rest of members to get site-wide results. Define
{\small
	\begin{align*}
	X_{rest} &=\text{ total value of the metric X for members NOT in the targeted segment}\\
	Y_{rest} &= \text{ total value of the metric } X^{1-\epsilon} \text{ for members NOT in said segment}\\
	n_{all} &= \text{ total number of members  (entire population)}\\
	n_{seg} &= \text{ number of targeted members (size of the targeted segment)}\\
	X_c &= \text{ total sum of the metric X for the control group}\\
	Y_c &= \text{ total sum of the metric } X^{1-\epsilon} \text{ for the control group}\\
	n_c &= \text{ number of people in control}\\
	X_t &= \text{ total sum of the metric X for the treatment group}\\
	Y_t &= \text{ total sum of the metric } X^{1-\epsilon} \text{ for the treatment group}\\
	n_t &= \text{ number of people in treatment}
	\end{align*}
}
Then we can estimate the site-wide values in the control universe using
\begin{align*}
X_c^{sw} &= X_{rest} + \frac{n_{seg}}{n_c} X_c\\
Y_c^{sw} &= Y_{rest} + \frac{n_{seg}}{n_c} Y_c\\
A_c^{sw} &= 1 - \frac{\big[\frac{Y_c^{sw}}{n_{all}}\big]^{1/(1-\epsilon)}}{\frac{X_c^{sw}}{n_{all}}}.
\end{align*}
Similarly for the treatment universe, we have
\begin{align*}
X_t^{sw} &= X_{rest} + \frac{n_{seg}}{n_t} X_t\\
Y_t^{sw} &= Y_{rest} + \frac{n_{seg}}{n_t} Y_t\\
A_t^{sw} &= 1 - \frac{\big[\frac{Y_t^{sw}}{n_{all}}\big]^{1/(1-\epsilon)}}{\frac{X_t^{sw}}{n_{all}}}.
\end{align*}
Finally, we can compare the site-wide inequality impact of treatment and control by comparing $A_t^{sw}$ to $A_c^{sw}$. The relevant asymptotic variance for that comparison is shown in section \ref{swiVar} of the appendix.

%

\section{Using this method at scale for business impact}

\subsection{A highly scalable implementation on Spark}

In order to fully leverage inequality A/B testing, we built a system that could handle the large scale of A/B testing at LinkedIn: thousands of concurrent experiments, millions of users, and dozens of metrics of interest. In this section, we outline the design of this system.

\subsubsection*{Single experiment} \label{singleexp}

In order to compute an Atkinson index on an arbitrarily large number of users, a distributed approach is needed. Thankfully, the Atkinson index can be decomposed as a function of sums of powers of metrics, making parallelization possible.  One can first raise each user's
metrics to various powers of $\epsilon$ and $1-\epsilon$, then sum across all
nodes, and compute the final index. To this end, we make available as simple Spark-SQL user-defined aggregation function (UDAF) for Apache Spark.
For simple use cases, it can be leveraged as follows: first, the aggregator needs to be initialized  \texttt{ val atkinsonAgg = new AtkinsonAggregator(epsilon) }; then it can simply be applied to the data as follows: \\ \texttt{  val inequalityStatistics = \\ dataFrame.select(atkinsonAgg(\$''metric''))}.

This UDAF can simply be copied or can be used as part of our open-source package.

\subsubsection*{Multi-experiment}

\begin{figure*}
	\includegraphics[width=\textwidth]{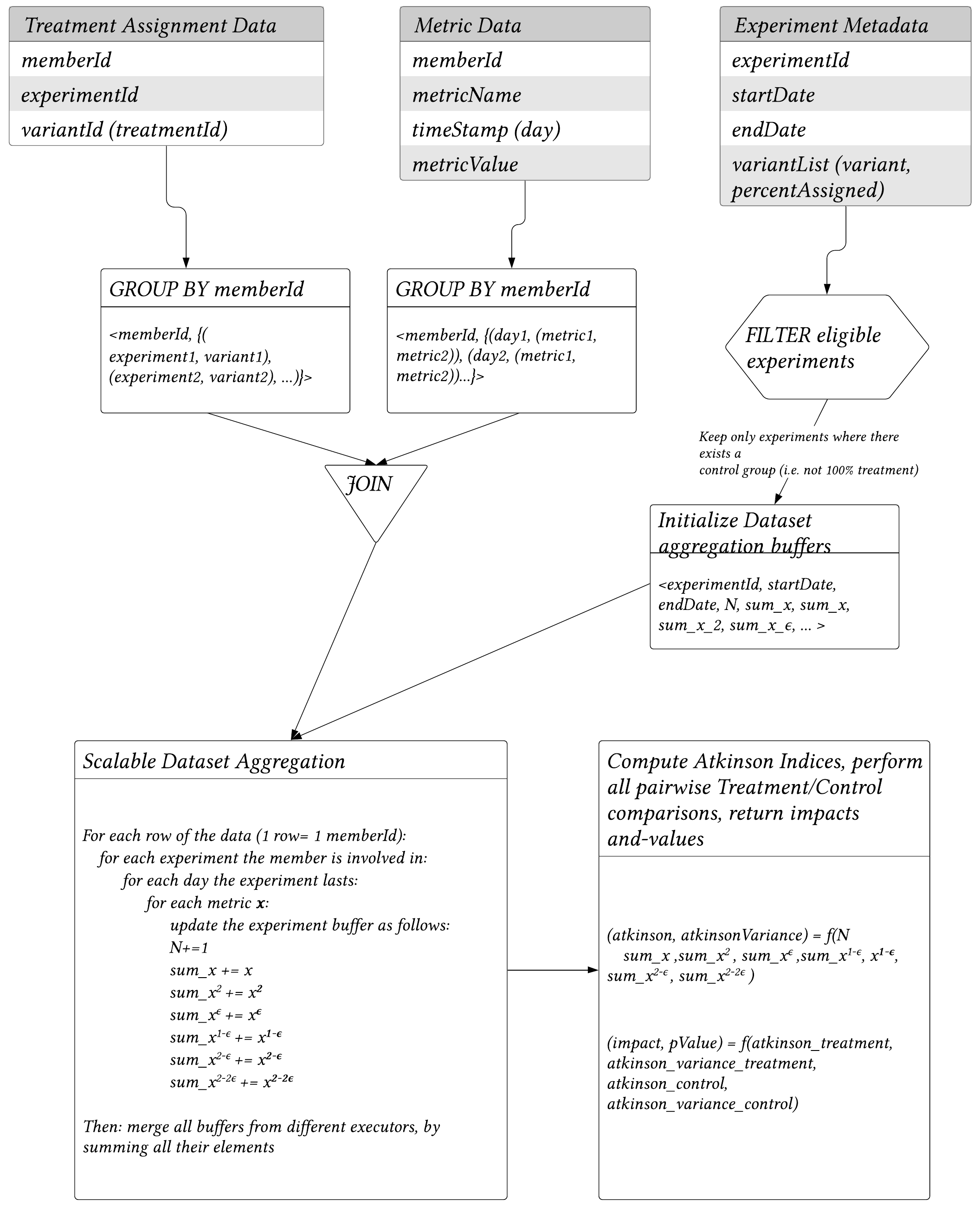}
	\caption{Flowchart of scalable inequality impact computation. This is a simplified view. }
	\label{multiExpflowChart}
	{\small
		In particular, it omits the fact that experiments are often split into population segments, and we also iterate over those. It also omits several sanity checks performed along the way, such as checking for Sample Size Ratio Mismatch (i.e. when an experiment is configured to have a certain percentage of the population treated, but significantly different percentage is recovered from the tracking data.) Finally, the computation of site-wide inequality impact follows a very similar logic, but is not shown for simplicity}
\end{figure*}

To perform large-scale A/B comparisons of inequality indices, we resorted to a scalable Spark Dataset aggregation approach. From a list of all currently running experiments and metadata, we initialize a large number of buffers. These buffers are designed to store all the primitives necessary for the computation of an Akinson index and of its variance. These buffers are then updated in parallel while reading the metric and treatment assignment data and merged. The buffers are finally converted to Atkinson indices and variances, and compared pairwise to perform statistical inference. We outline the multi-experiment algorithm as a flowchart in Figure \ref{multiExpflowChart}, and also make it available open-source.

\subsection{Organizational review process}

Computing the inequality impact of all experiments is, of course, just the beginning. In order to have an impact on the user experience, we also outlined an organizational process, based on two primary mechanisms. First, we rely on positive recognition and reinforcement of initiatives that reduce inequality in core metrics. Second, we propose an alert system for initiatives (experiments) that increase inequality, especially when such increases might be unintentional or avoidable.

\begin{figure}
	\begin{center}
		\includegraphics[width=0.7\textwidth]{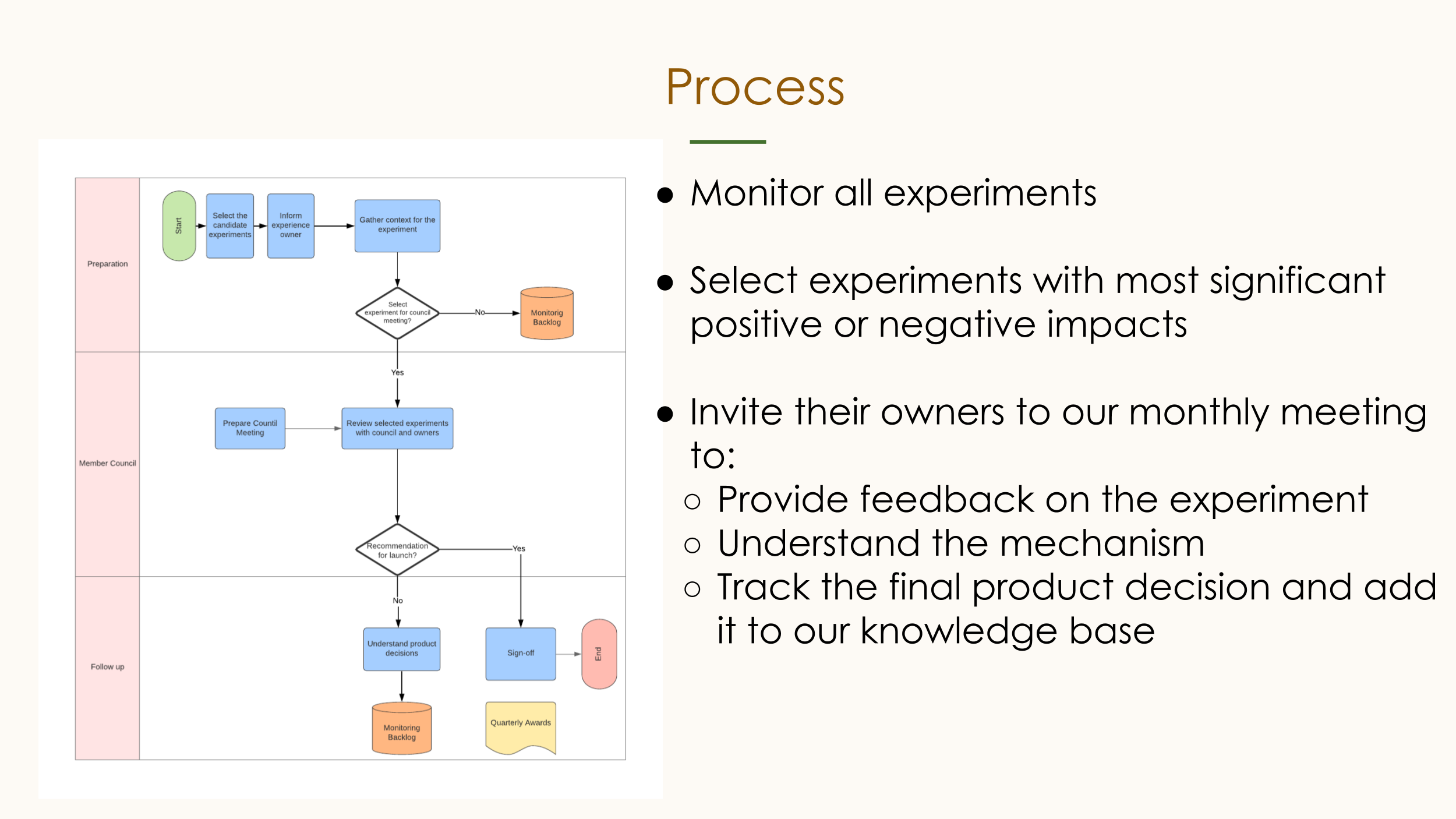}
		\caption{The LinkedIn process to review experiments flagged by this method}
		\label{process}
	\end{center}
\end{figure}

Every month, we analyze all experiments that have been run. As outlined above, the main pieces are a large-scale Spark job, and a simple UI. We then select the most impactful experiments -- the ones that had the highest absolute value inequality impact --  for further review. A core team reaches out to experiment owners to get as much context as possible about the exact nature of the intervention tested in the experiment, and about its primary intent. A monthly meeting is then set up with the experiment owners, where we share the inequality result with them, learn from them whether it was intended or not, and decide on next steps. We save all the results and relevant context, in order to build an extensive knowledge base about which types of interventions increase or decrease inequality between our members. There can be many follow-ups, depending on the finding: if the experiment reduces inequality, the owners are typically encouraged to hear it. They often invest more in the same direction, as was the case for our leading example in this paper, described in section \ref{instajobs}. If the result is an increase in inequality, we typically recommend a deep dive to understand the causes. This often entails trying to discover if the inequality is being created along a particular dimension, i.e., between discernible groups of users: for example, users of different countries, or users with different network strength (see \ref{social_capital}). Our review process is illustrated in Figure \ref{process}.

\section{Example of an experiment: Instant job notifications} \label{instajobs}
A new feature on LinkedIn, named Instant Job Notifications, informs active job seekers of new job postings. Within a few minutes of the time a new job is posted, Instant Job Notifications uses the Automated Sourcing algorithm to find seekers for whom this job is a good fit, and sends them a push and an in-app notification. Our analysis shows that
it both decreased inequality in engagement and job applications, but
also reduced the gap between job applications and engagement of
well-connected members versus poorly connected members. For an explanation of how we distinguish well-connected member, see Appendix \ref{social_capital}. This is for two main
reasons: first, members with poorer networks were less likely to be
referred to these jobs or to hear about them through their network.
Second, members tend to self-censor if they see that a job opening has already received applications, and this is more likely to be true for particular group of users, like women. Sharing the job opportunity before it had received a high number of applicants made users more likely to apply.

\begin{figure}
	\begin{center}
		\includegraphics[width=0.7\textwidth]{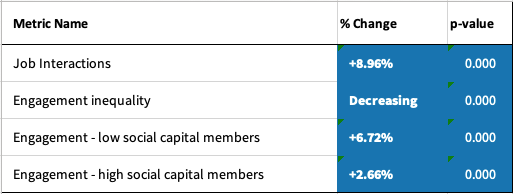}
		\caption{Impact of Instant Job Notifications, multiplied by a $\chi_1^2$ distributed random number for confidentiality.} \label{instaJob}
	\end{center}
\end{figure}

The feature increased the number of job applications and the chance of the application receiving interaction from the employer. More importantly for us, from the job posters perspective, this featured decreased inequality in the number of applications received between job postings. That is, the jobs that previously received very few applications started receiving more. Figure \ref{instaJob} shows the impact of Instant Job Notifications on some of the relevant metrics. The numbers are multiplied by a $\chi_1^2$ distributed random number.

\section{Analyzing thousands of experiments: what we learned so far}
Fairness through experimentation should not only be about setting guardrails with A/B testing and detecting experiments that could be potentially damaging to our members. It is also about finding inspiration to create a more inclusive product. Contrary to ``top-down'' fairness criteria or ex-ante requirement, we try to learn from the thousands of colleagues and engineers as they experiment on the features they are building. Through this, we are compiling a knowledge base of types of interventions that seem to reduce inequality, so that it may be used as a guide for development of our next products. In the last section of this paper, we share a few of the things we have been able to learn so far.

\subsection{Metric-neutral interventions are often not neutral for everyone}

In many situations, teams may try to implement ``neutral''  interventions: for example, when performing a back-end infrastructure change, or when trying to ``boost'' or promote a specific product on the platform while making sure no other product suffers. The principal method of monitoring this neutrality is looking at a treatment-control comparison of the average impact (often called lift) in the experimentation platform. However, even if no average impact is detected, there is often an inequality impact. In other words, metrics are not affected on average, but some users are. This makes inequality a critical aspect to monitor in such experiments, as ``neutral'' should mean neutral for everyone. We have seen examples of this while trying to promote a specific feature: it had no negative impact on other elements of the site on average, but increased engagement inequality, and was eventually canceled.

\subsection{Notifications are a powerful tool}
Notifications have a strong impact on inequality of engagement, as they are a powerful tool to bring the least-engaged members to the site. Similarly, reducing notifications to the most engaged members reduces engagement inequality.

\subsubsection*{New Member onboarding}

Making sure more people benefit from LinkedIn requires adding new members, and helping them familiarize themselves with the platform and its value proposition to them. To that end, the quality of the onboarding process is of paramount importance, as new members have a high chance of dropping off a new service or platform in the first few days. We analyzed an experiment that tried to accompany new, onboarding members better, through the use of nudges and notifications. Before that experiment, many new members did not receive any notifications. In this experiment, new members now received 1-2 push notifications
during their first week - encouraging them to take key actions to build their network and create value for themselves. This had a positive impact on average engagement, but also on inequality of engagement, since it primarily helped members who were at the highest risk of dropping off the site.

\subsubsection*{Notification batching for highly engaged members}
For optimal user experience, it may be necessary to batch notifications together. For example, if a member receives 55 actions on their post, we can either send her 55 notifications, or bundle them together in progressively: send one for the first like, then another one for ten likes, and so on. We analyzed a similar experiment and saw that notification batching reduced inequality, as it primarily affected the most engaged members.

\subsection{Site speed and availability matters to inclusiveness}
We found that many interventions relating to site speed and reliability had a disproportionately positive impact on the least engaged members, and reduced inequality. This makes sense, as those may be members with slower devices and connections.

\subsubsection*{The low-bandwidth LinkedIn app}
Several experiments on the low-bandwidth LinkedIn app showed a strong positive impact, both on average engagement and on inequality in engagement. The low-bandwidth LinkedIn app targets members with slower network speeds or slower devices. Enabling that feature, or adding features that brought the low-bandwidth experience closer to the main experience (like enabling hashtags, for example), had positive inclusiveness effects. 


\subsection{On social network platforms, social capital matters for inclusiveness}
Once an inequality impact (positive or negative) is found, we seek to understand it, in particular by asking whether we can identify two or more groups that are being affected deferentially by the experiment. Note that we care about inequality whether or not it can be summarized as a differential impact on different groups, but identifying groups helps with interpretation. We have repeatedly found that a member's network strength (i.e., her social capital, how well-connected she is) often has an impact on how much value she can get out of a social network.

\subsection{Both positive and negative inequality impacts are often unintended}
Throughout over a year of experiment review meetings and learning from experiment owners, we often found that the inequality impact we surfaced was unintended. On occasion, it was positive, and on others, negative. Designers may often think about users in an idealized fashion: as a representative, average user that does not actually exist. This may pose inclusiveness challenges, as it runs the risk of leaving users who do not resemble the idealized average behind. 

\section{Conclusion}

We have shared a new approach to responsible design, complementary to
mainstream approaches to algorithmic fairness. It emphasizes inequality
measurement in experimentation, to focus on final impacts of
algorithms (including user's behavioral responses), rather than on the
properties of the algorithms only. It is highly scalable and deployed
throughout LinkedIn. It has allowed us to identify and reinforce
inequality-reducing initiative and is a valuable way for companies to
monitor potentially unintended consequences of their experiments.

\nocite{*}
\bibliographystyle{plain}
\bibliography{bib}

\appendix
\section{Appendix}
\spacing{1}

\subsection{Variance of the Atkinson index}
\begin{proof}{Proof of Theorem 1}
Note that we can write
\begin{align*}
A_\epsilon(x_1 , \ldots , x _n) = f (\frac{1}{n}\sum_{i=1}^n x_i^{1-\epsilon},  \frac{1}{n}\sum_{i=1}^n x_i),
\end{align*}
for $ f(x , y)  = 1- \frac{x^{\frac{1}{1-\epsilon}}}{y}$. We will use the CLT for $(\frac{1}{n}\sum_{i=1}^n x_i^{1-\epsilon},  \frac{1}{n}\sum_{i=1}^n x_i)$ along with the delta method to find the asymptotic distribution of $A_\epsilon(x_1 , \ldots , x _n) $.
Let $\mu_\alpha$ denote the population mean of $x^{\alpha}$. Then, the CLT asserts
\begin{align*}
\sqrt{n} \left( (\frac{1}{n}\sum_{i=1}^n x_i^{1-\epsilon},  \frac{1}{n}\sum_{i=1}^n x_i) - (\mu_{ 1 -\epsilon} ,  \mu_1) \right)  \rightarrow \mathcal{N}\left( (0,0) , \Sigma \right),
\end{align*}
where the limiting covariance matrix can be consistently estimated as
\begin{align*}
\widehat{\Sigma}_{11} &= \frac{1}{n}\sum_{i=1}^n x_i^{2-2\epsilon} - (\frac{1}{n}\sum_{i=1}^n x_i^{1-\epsilon})^2, \\
 \widehat{\Sigma}_{21} &=  \widehat{\Sigma}_{12} = \frac{1}{n}\sum_{i=1}^n x_i^{2-\epsilon} -(\frac{1}{n}\sum_{i=1}^n x_i)\times (\frac{1}{n}\sum_{i=1}^n x_i^{1-\epsilon}),\\
  \widehat{\Sigma}_{22} &=  \frac{1}{n}\sum_{i=1}^n x_i^{2} - (\frac{1}{n}\sum_{i=1}^n x_i)^2.
\end{align*}
The delta method asserts that
\begin{align*}
\sqrt{n} \left( f (\frac{1}{n}\sum_{i=1}^n x_i^{1-\epsilon},  \frac{1}{n}\sum_{i=1}^n x_i) - f(\mu_{ 1- \epsilon}, \mu_1)  \right) \rightarrow \mathcal{N} (0, \sigma^2),
\end{align*}
where $\sigma^2 = (\nabla  f(\mu_{ 1- \epsilon}, \mu_1) )^T \Sigma ( \nabla  f(\mu_{ 1- \epsilon}, \mu_1)).$ We can use sample plug-in estimates of $\Sigma$, $\mu_1$, and $\mu_{1 - \epsilon}$ to obtain an estimate of $\sigma$, as
\begin{equation}
\begin{split}\label{Atk:var}
\sigma_n^2 = \widehat{\Sigma}_{11} \times\frac{(\frac{1}{n}\sum_{i=1}^n x_i^{1-\epsilon})^\frac{2 \epsilon}{1-\epsilon}}{(1-\epsilon)^2\left(\frac{1}{n}\sum_{i=1}^n x_i \right)^2 } - 2 \widehat{\Sigma}_{12} \times  \frac{(\frac{1}{n}\sum_{i=1}^n x_i^{1-\epsilon})^\frac{1 + \epsilon}{1-\epsilon}}{(1-\epsilon)\left(\frac{1}{n}\sum_{i=1}^n x_i \right)^3 } +  \\ \widehat{\Sigma}_{22} \times \frac{(\frac{1}{n}\sum_{i=1}^n x_i^{1-\epsilon})^\frac{2}{1-\epsilon}}{\left(\frac{1}{n}\sum_{i=1}^n x_i \right)^4 }.
\end{split}
\end{equation}
Law of large numbers guarantees that $\sigma_n$ is a consistent estimator of $\sigma$. This along with Slutsky's theorem proves the result.
\end{proof}

\subsection{Variance for Site-wide impact} \label{swiVar}
The variance of the SWA can be estimated similarly. In fact, it is given by the same formula as \ref{Atk:var}, except that the sample averages are replaced with sitewide extrapolation defined in section \ref{swi-atk}, and the variance is scaled appropriately to reflect the relative size of the eligible population compared to the entire member-base.
{\tiny 
\begin{equation}
\begin{split}
\sigma_n^2 = (\frac{n_{seg}}{n_{all}})^2   \times \\  \left[\widehat{\Sigma}_{11} \times\frac{(Y^{sw}/n_{all})^\frac{2 \epsilon}{1-\epsilon}}{(1-\epsilon)^2\left(X^{sw}/n_{all} \right)^2 } - 2 \widehat{\Sigma}_{12} \times  \frac{(Y^{sw}/n_{all})^\frac{1 +  \epsilon}{1-\epsilon}}{(1-\epsilon)\left(X^{sw}/n_{all} \right)^3 } +  \widehat{\Sigma}_{22} \times \frac{(Y^{sw}/n_{all})^\frac{2}{1-\epsilon}}{\left(X^{sw}/n_{all} \right)^4 }\right].
\end{split}
\end{equation}
}


\subsection{An important demographic cut: network strength} \label{social_capital}

Once we know an experiment impacts inequality, it is natural to look for the underlying reason. One standard approach is to compare the impact of the experiment on different cohorts of members. We take this approach in this section with a focus on member attributes that describe members' network on LinkedIn. There are various ways to extract member features from networks.
Motivated by  \cite{watts1998collective}, \cite{Saint-Jacques2018dissertation} \cite{Granovetter1973}, \cite{Granovetter1995}, on the power of weak ties we use structural network diversity, defined below, as a proxy for social capital.  \cite{Saint-Jacques2018dissertation}  establishes that structural network diversity has a measurable positive on members' career outcomes. In particular, when the structural network diversity increases, people are more likely to change jobs. Structural network diversity is defined through the local clustering coefficient, introduced by \cite{watts1998collective}, and is defined as follows.

A graph \(G = (V,E)\) formally consists of a set of vertices\(V\) (all LinkedIn members) and a set of edges \(E\) between them (e.g. all the connections between members). An edge \(e_{ij}\) connects vertex \(v_i\) with vertex \(v_j\).
The neighborhood of vertex \(i\) is defined as \(N_i = \{ v_j; e_{ij} \in E\}.\) In our case, for a specific member \(i\), this is the set of all her connections. The number of connections is given by \(k_i = |N_i|\).
Local clustering coefficient for vertex \(v_i\) is the number of connections between the neighbors of \(v_i\), i.e. members of \(N_i\), divided by the total number of such possible connection. That is mathematically equal to
\[d_i = \frac{2|\{ e_{jk} ;\ v_k, v_j \in N_i , e_{jk} \in E\}|}{k_i (k_i -1)}.\]
Structural network diversity is defined as \(1 - d_i\).

We divide members into three buckets based on the structural network diversity score, the bottom 20\%  are the members with low social capital and the top 20\% are the members with high social capital.

\clearpage
\subsection{Scalable computation of the Atkinson Index and its variance Using Spark-SQL}
We share the simple single-metric implementation of the Atkinson Index computation here, for illustrative purposes. This code is directly usable by any spark user as outlined in section \ref{singleexp}. The code for the multi-experiment, multi-metric scalable aggregation is significantly too large to fit in this appendix, but is available upon request as an open-source project. This code is written for legibility, rather than low-level optimization (we could make fewer calls calls to Math.pow(), for example.)
 \label{udaf}
{\tiny
\begin{lstlisting}[backgroundcolor = \color{light-gray}]

import org.apache.spark.sql.expressions.MutableAggregationBuffer
import org.apache.spark.sql.Row
import org.apache.spark.sql.types._

/** a Spark User-Defined aggregation function (UDAF)
  * to compute the Atkinson Index and its variance
  *
  * @param e Atkinson's epsilon, measuring inequality aversion parameter
  * @author Amir Sepehri (math), Guillaume Saint-Jacques (scalable implementation)
  */
class AtkinsonAggregator(val e: Double) extends
  org.apache.spark.sql.expressions.UserDefinedAggregateFunction {

  /** input schema of UDAF */
  override def inputSchema: org.apache.spark.sql.types.StructType =
    StructType(StructField("value", DoubleType) :: Nil)

  /** buffer internal fields :
    * N : number of observations
    * S : sum of observations
    * S1me : sum of observations to the power of 1-epsilon
    * S2m2e : sum of observations to the power of 2 - 2 epsilon
    * S2me : sum of observations to the power of 2-epsilon
    * S2 : sum of observations to the power of 2
    */
  override def bufferSchema: StructType = StructType(
    StructField("N", LongType) ::
      StructField("S", DoubleType) ::
      StructField("S1me", DoubleType) ::
      StructField("S2m2e", DoubleType) ::
      StructField("S2me", DoubleType) ::
      StructField("S2", DoubleType) :: Nil)

  /** output Schema:
    * - N is the number of observation
    * - atkinsonIndex is the Atkinson Index
    * - sigmaSq is the estimated variance of the atkiosn Index
    */
  override def dataType: DataType = StructType(
    StructField("N", DoubleType) ::
      StructField("atkinsonIndex", DoubleType) ::
      StructField("sigmaSq", DoubleType) :: Nil)

  /** this is a deterministic algorithm */
  override def deterministic: Boolean = true

  /** all accumulators initialized at zero */
  override def initialize(buffer: MutableAggregationBuffer): Unit = {
    buffer(0) = 0L
    buffer(1) = 0.0
    buffer(2) = 0.0
    buffer(3) = 0.0
    buffer(4) = 0.0
    buffer(5) = 0.0
  }

  /** accumulator update function when presented with new data */
  override def update(buffer: MutableAggregationBuffer, input: Row): Unit = {
    val x = input.getAs[Double](0)
    buffer(0) = buffer.getAs[Long](0) + 1
    buffer(1) = buffer.getAs[Double](1) + x
    buffer(2) = buffer.getAs[Double](2) + Math.pow(x, 1 - e)
    buffer(3) = buffer.getAs[Double](3) + Math.pow(x, 2 - 2 * e)
    buffer(4) = buffer.getAs[Double](4) + Math.pow(x, 2 - e)
    buffer(5) = buffer.getAs[Double](5) + Math.pow(x, 2)
  }

  /** buffer merge function */
  override def merge(buffer1: MutableAggregationBuffer, buffer2: Row): Unit = {
    buffer1(0) = buffer1.getAs[Long](0) + buffer2.getAs[Long](0)
    buffer1(1) = buffer1.getAs[Double](1) + buffer2.getAs[Double](1)
    buffer1(2) = buffer1.getAs[Double](2) + buffer2.getAs[Double](2)
    buffer1(3) = buffer1.getAs[Double](3) + buffer2.getAs[Double](3)
    buffer1(4) = buffer1.getAs[Double](4) + buffer2.getAs[Double](4)
    buffer1(5) = buffer1.getAs[Double](5) + buffer2.getAs[Double](5)
  }

  /** when accumulation is over, compute the actual index and variance
    *
    * @param buffer the accumulators
    * @return three terms: number of observations, atkinson Index, variance of index
    */
  override def evaluate(buffer: Row): Any = {
    /** number of observations */
    val N = buffer.getAs[Long](0).toDouble
    /** average */
    val S = buffer.getAs[Double](1) / N
    /** average of terms raised to the power of 1 - epsilon */
    val S1me = buffer.getAs[Double](2) / N
    /** average of terms raised to the power of 2 - 2 epsilon */
    val S2m2e = buffer.getAs[Double](3) / N
    /** average of terms raised to the power of 2-epsilon */
    val S2me = buffer.getAs[Double](4) / N
    /** average of terms raised to the power of 2 */
    val S2 = buffer.getAs[Double](5) / N
    /** actual Atkinson Index */
    val atkinsonIndex = 1 - (Math.pow(S1me, 1 / (1 - e)) / S)
    /** first term of var-covar matrix */
    val S11 = S2m2e - Math.pow(S1me, 2)
    /** second term of var-covar matrix */
    val S12 = S2me - S * S1me
    /** third term of var-covar matrix */
    val S22 = S2 - Math.pow(S, 2)
    /** final estimated variance of the Atkinson Index */
    val sigmaSq = (
      S11 * Math.pow(S1me, 2 * e / (1 - e)) / (Math.pow(1 - e, 2) * Math.pow(S, 2))
        - 2 * S12 * Math.pow(S1me, (1 + e) / (1 - e)) / ((1 - e) * Math.pow(S, 3))
        + S22 * Math.pow(S1me, 2 / (1 - e)) / Math.pow(S, 4)) / N
    (N, atkinsonIndex, sigmaSq)
  }
}

\end{lstlisting}

}

\end{document}